\documentclass[%
 reprint,
 amsmath,amssymb,
 aps]{revtex4-2}

\usepackage{graphicx}
\usepackage{dcolumn}
\usepackage{bm}
\usepackage[mathlines]{lineno}
\usepackage{enumitem}
\usepackage[utf8]{inputenc}
\usepackage[T1]{fontenc}
\usepackage{mathptmx}
\usepackage{etoolbox}
\usepackage{hyperref}
\usepackage{setspace}

\hypersetup{
	colorlinks = true,
	linkcolor = blue,
	anchorcolor = blue,
	citecolor = blue,
	filecolor = blue,
	urlcolor = blue
}

\makeatletter
\def\@email#1#2{%
 \endgroup
 \patchcmd{\titleblock@produce}
  {\frontmatter@RRAPformat}
  {\frontmatter@RRAPformat{\produce@RRAP{*#1\href{mailto:#2}{#2}}}\frontmatter@RRAPformat}
  {}{}
}%
\makeatother

\begin{document}

\preprint{APS/123-QED}

\title{Microscale architected materials for elastic wave guiding: Fabrication and dynamic characterization across length and time scales}
\author{Vignesh Kannan}
\email{vignesh.kannan@polytechnique.edu}
\affiliation{Laboratoire de M{\'e}canique des Solides, {\'{E}}cole Polytechnique, Palaiseau, 91128, France}
\altaffiliation{Mechanics \& Materials Laboratory, ETH Zurich, Zurich, 8092, Switzerland.}
 
\author{Charles Dorn}%
 \email{cdorn@uw.edu}
\affiliation{Department of Aeronautics and Astronautics, University of Washington, Seattle, WA, 98195, USA.}%
\altaffiliation{Mechanics \& Materials Laboratory, ETH Zurich, Zurich, 8092, Switzerland.}

\author{Ute Drechsler}
\affiliation{%
IBM Research - Zurich, Ruschlikon, 8803, Switzerland.
}%

\author{Dennis M.~Kochmann}
\email{dmk@ethz.ch}
\affiliation{%
Mechanics \& Materials Laboratory, ETH Zurich, Zurich, 8092, Switzerland.
}%

\date{\today}

\begin{abstract}
We present an experimental protocol for the fabrication and characterization of scalable microarchitected elastic waveguides. Using silicon microfabrication techniques, we develop free-standing 2D truss-based architected waveguides with a maximum diameter of 80~mm, unit cells size of $100$~\textmu m, and minimum beam width of $5$~\textmu m, thus achieving scale separation. To characterize elastic wave propagation, we introduce a custom-built scanning optical pump-probe experiment that enables contactless excitation of elastic wave modes and full spatio-temporal reconstruction of wave propagation across hundreds of unit cells with sub-unit cell resolution. Results on periodic architectures show excellent agreement with finite element simulations and equivalent experimental data at larger length scales. Motivated by scalable computational inverse design, we fabricate a specific example of a spatially graded waveguide and demonstrate its ability to guide elastic waves along an arbitrary pre-designed path.
\end{abstract}

\keywords{metamaterials, elastic wave guiding, silicon microfabrication, heterodyne interferometry}
\maketitle


\section{Introduction \label{sec:intro}}
Architected metamaterials present an unprecedented design space to control material functionality across length and time scales. The control of elastic wave propagation or ``wave guiding'' is a particularly exciting, albeit challenging problem. A fundamental understanding of wave propagation through periodic architectures, originating from solid-state physics \cite{brillouin1946wave}, has pervaded the area of photonic metamaterials and lead to contemporary research on phononic/elastic wave guiding through structural networks \cite{hussein2014dynamics,phani2006wave}. This design space of architectures for engineered wave motion is vast, with fundamental functional units ranging from grains and grain boundaries \cite{Thompson} to phases in composites \cite{Vasseur1998,Sigmund2003} to beams, plates, and shells \cite{trainiti2016wave,zelhofer2017acoustic,bilal2011ultrawide}. Structural metamaterials such as beam networks are particularly attractive for elastic wave guide design because of our ability to reliably close the loop between fabrication, experimental characterization, and modeling of the dynamic response. 

Earlier studies in the design of metamaterial wave guides relied on the periodicity of the architecture, showing fascinating opportunities to control the directionality and frequency selection of elastic waves \cite{hussein2014dynamics,phani2006wave}. Advancements and commercialization of 3D printing technology, along with non-contact laser-based measurement capabilities, have enabled the realization of these phenomena in experiments \cite{Schaeffer2017,Miniaci2018,telgen2024rainbow}. Many of those studies suffer from the limitation to polymer-based base materials for fabrication, whose viscoelastic damping makes it challenge to associate wave attenuation to either base material-intrinsic dissipation or structural wave guiding mechanisms and to decouple these material vs.\ structural phenomena. This bottleneck can be overcome by traditional manufacturing techniques such as laser or water jet cutting of metallic plates \cite{dorn2023a,telgen2024rainbow}. However, those are severely limited in resolution and cannot achieve such high densities of unit cells within reasonable sample sizes. 

More recently, spatially varying architectures have significantly enlarged the design space for elastic wave guiding \cite{Xie2018,Zhao2020,Chaplain2020}. Due to limitations in classical analysis based on Bloch wave theory, modeling elastic waves through such spatially graded architectures necessitated the development of new tools \cite{Craster2010,schnitzer2017waves}. Direct finite element-based dynamic simulations are accurate, but the high temporal and spatial resolution required makes them prohibitively expensive to extract short-wavelength information across a large spatial domain of varying architecture. By contrast, recent developments using ray theory, based on the assumption of local periodicity in a smooth structural grading \cite{Dorn2022,dorn2023a}, have demonstrated the ability to capture short-wavelength elastic waves in spatially graded networks, without explicitly resolving the underlying architecture. Ray tracing further provides an efficient modeling tool that enables the inverse design of complex spatial gradings to customize wave guiding capabilities \cite{Dorn2023b}. 

These advances in computational modeling and design pose new challenges for experimental mechanics. Inverse design can generate spatially-varying, locally-periodic architectures that involve hundreds of thousands of unit cells \cite{DornKannan2025}. Fabricating samples with such a large density of unit cells is a first challenge. State-of-the-art 3D printing technology may achieve high spatial resolution but is primarily limited to viscoelastic, polymer-base materials \cite{meza2017reexamining,Vyatskikh2018,bauer2017nanolattices,shaikeea2022toughness,krodel2016microlattice,kiefer2024multi,harinarayana2021two} with significant base material damping. Despite recent efforts to print metallic metamaterials at small scales \cite{Saccone2022}, the trade-off between spatial resolution and size of the structure will continue to limit the use of such techniques for wave guiding. 

To overcome the aforementioned limitations, we employ techniques from the microelectronics industry, using silicon-based microfabrication. Microfabrication has served as a powerful tool in experimental mechanics of materials, e.g., probing size effects in material behavior \cite{Espinosa2003,Uchic2004,Zhu2007,Naceri2025}, fabricating devices for biomechanics \cite{Loh2007,Hosmane2011}, enabling high-throughput experiments \cite{Boyce2009,Kai2023}, and ensuring the durability of micro-electromechanical systems (MEMS) and nano-electromechanical systems (NEMS) \cite{Chasiotis2004,Sharpe2007}. In many of those applications, as in microelectronics, the ``device'' is fabricated on a silicon (Si) wafer, with multiple ``devices'' on a single wafer. In the context of wave guiding, this limits the modes that could otherwise be excited in a ``free-standing device with internal architecture''. Hence, the wafer becomes the ``device'', and its functional units, i.e., the unit cells, are its ``components''. The realization of such waveguides for experimental studies has been limited. Cha et al.~\cite{Cha2018a,Cha2018b} developed a topological nano-electromechanical metamaterial (NEMM) involving a pattern of free-standing SiN nanomembranes on a doped Si substrate, with target frequencies on the order of 10~MHz. Thelen et al.~\cite{Thelen2021} studied elastic wave propagation and dispersion in nanoporous Si wafers. Acoustic wave transmission was investigated in suspended SiO$_2$ architected membranes in \cite{olsson2008a,olsson2008b}. Recently, one-dimensional microscale arrays have been reported for sensing and energy harvesting applications \cite{maspero2023phononic, de2024localized}. Such microfabrication techniques are excellently suited to also enable the scalable creation of spatially graded elastic wave guides, yet there seem to be no studies on the fabrication and characterization of micro-architected free-standing metamaterial waveguides with unit cell densities on the order of $10^{3} - 10^{4}$ per mm$^{2}$ --- which are critical to the realization of high-resolution spatially graded elastic waveguides, now accessible through efficient computational design approaches. 

Characterizing the dynamic response of such thin-film devices poses the second challenge and exceeds the capabilities of conventional testing techniques at macroscopic scales. Excitation of an elastic wave, conventionally achieved by contact-based piezoelectric transducers, becomes challenging due to size and frequency limitations -- the process would require the synthesis of micro-scale transducers. (And even if those could be produced, they would limit the spatial distribution of initial conditions and the accessible frequency range.) Interdigital transducers (IDTs) \cite{White1965}, primarily used for surface acoustic wave studies, are infeasible due to limited control of the loading profile and the complexity of their fabrication on free-standing films. Photocoustic excitation by pulsed laser sources \cite{Davies1993,Krishnaswamy2008} has been used as a powerful tool in the study of the dynamic properties of thin films \cite{Nelson1982,Rogers2000}, crystals \cite{Stoklasov2021}, defects in materials \cite{RezazadehKalehbasti2019}, and surface acoustic wave guides \cite{Khanolkar2015,Hiraiwa2016}. Moreover, the measurement of propagating waves in such small structures is inaccessible to most commercial laser Doppler vibrometers, and the few that may be used involve a large cost-to-value ratio and limited customizability. This motivates the in-house development of scanning interferometers with high spatial, displacement, and temporal resolutions \cite{Dewhurst1999,Sottos1991,Otsuka2018}. 

This paper presents a novel fabrication protocol to generate $80$~mm diameter free-standing elastic waveguides with unit cell densities on the order of $10^{3} - 10^{4}$ unit cells per mm$^{2}$ (Sec.~\ref{subsec:fabrication}), followed by a home-built scanning photoacoustic pump-probe experiment to excite and measure elastic waves across a wide frequency range of tens of kHz up to $10$~MHz and current spatial and displacement resolutions on the order of \textmu m and nm respectively (Sec.~\ref{subsec:photacoustic}). This fabrication protocol is tested on periodic architected films, whose comparison to finite element results and prior experimental measurements at macroscopic length scales confirms the validity (Sect.~\ref{subsec:periodic}). We finally present a realization of a spatially-graded elastic waveguide (Sec.~\ref{subsec:spatialgrading}), hence demonstrating the closed loop between high-resolution fabrication, experimental characterization, and scalable computation.

\section{Methods \label{sec:methods}}
\subsection{Fabrication of free-standing Si-based metamaterials\label{subsec:fabrication}}
A major challenge in multiscale mechanical design is the ability to separate the length scales of the small-scale architectural features and the macroscopic response of the (meta-)material. We employ the state of the art in microfabrication to overcome related challenges in fabricating waveguides (described in Sec.~\ref{sec:intro}). The main design objectives for fabrication are: 
\begin{itemize}[leftmargin=*]
    \item to generate a free-standing architected film, whose (in-plane) dimensions are orders of magnitude larger than the unit cell.
    \item to enable multi-point excitation and measurement of elastic waves from both sides of the film.
\end{itemize}

\begin{figure*}
    \includegraphics[width=\linewidth,keepaspectratio]{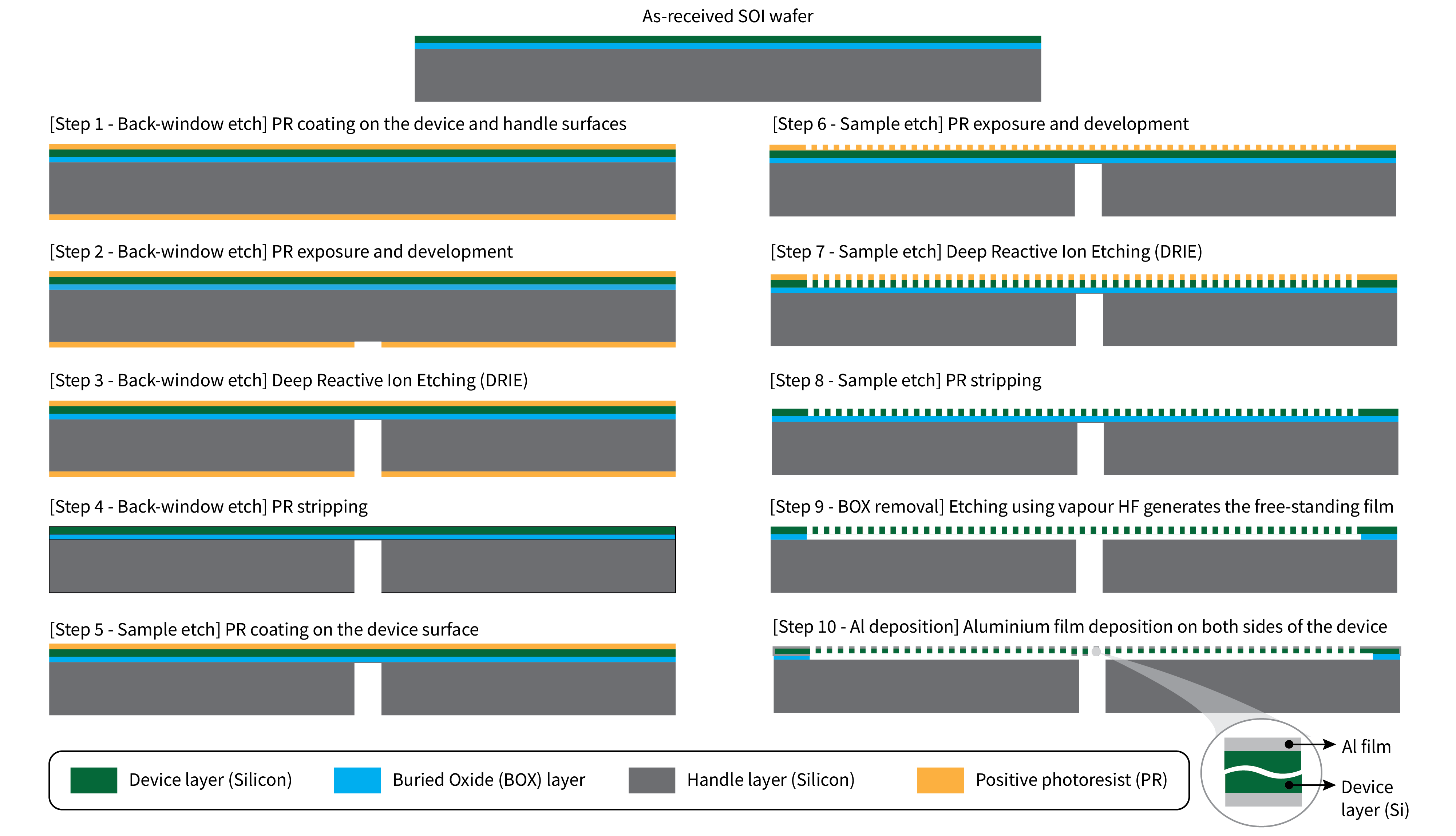}
    \caption{Schematic description of the microfabrication protocol. Fabrication was performed on as-received SOI wafers. Color references are indicated at the bottom of the figure. The inset in Step 10 shows the aluminum film deposited on the free-standing waveguide for optimal reflectivity during pump-probe characterization (Sec.~\ref{subsec:photacoustic}).}
    \label{fig:fabschem}
\end{figure*}

Samples were fabricated using commercially-processed 100~mm diameter Silion-On-Insulator (SOI) wafers. SOI wafers have three layers bonded to each other called the (1)~device, (2)~buried oxide (BOX), and (3)~handle layers (see the first schematic in Fig.~\ref{fig:fabschem}). Unit cells were etched on the device layer and the BOX layer removed to generate the free-standing architected film. The sample remains attached to the handle layer at the rim, enabling easy handling. Throughout this study, we use SOI wafers with the two different device layer thicknesses, $10$ and $15$~\textmu m, each with a tolerance of $\pm 0.5$~\textmu m. All our wafers have a BOX layer of thickness $2.4 \pm 0.12$~\textmu m, and handle layer $400 \pm 10$~\textmu m. The steps of the fabrication protocol are summarized as follows: 
\begin{enumerate}[leftmargin=*]
    \item The first sequence of steps etches a back-window in the handle layer for excitation of the elastic wave (details in Sec.~\ref{subsec:photacoustic}). Surfaces were first cleaned and subsequently treated with hexamethyl disilazone (HMDS) to promote adhesion of photo-resist layers. Two layers of positive photo-resist (PR) were deposited on the device layer (AZ4533 of $\sim 3$~\textmu m thickness) and on handle layers (AZ4562 of $\sim 6$~\textmu m). Both layers were baked at 120$^\circ$C for 2 minutes, each. The PR film on the device layer minimizes contamination of the surface during processing of the handle layer. 
    \item A photo-mask --- designed and developed in house with a direct laser writer --- was used to expose the handle layer, using a commercial (Karl Suss MA6) mask aligner. The exposed PR was developed using standard protocol (KOH solution -- AZ400K) in a spray developer, thus exposing the region to be etched on the handle layer. The exposed region was etched using Deep Reactive Ion Etching (DRIE), for $60 - 90$~minutes. 
    \item The PR films on both the device and handle layers were removed by rinsing with acetone and isopropyl alcohol, followed by immersion in dimethyl sulfoxide (DMSO) at 120$^\circ$C for about 10~minutes.
    \item To etch the waveguide architecture, a film of PR was re-deposited on the device layer. 
    \item Using a custom-developed photo-mask, the same processes described in step 2 were used to etch the elastic waveguide architecture on the device layer photo-resist and then developed for DRIE. 
    \item The architecture was etched into the device layer, using DRIE for about 15~minutes. The PR layer was then carefully stripped from the sample, using the same procedure outlined in step~3.
    \item The wafer was subjected to vapor hydrofluoric acid (HF) etching to remove the intermediate BOX layer, thus generating a free-standing $10$ or $15$~\textmu m thick micro-architected elastic waveguide. 
    \item As a final step, thin aluminum films of thicknesses $20 - 50$~nm were deposited on the front and back sides of the sample to act as transducers during optically-induced elastic wave excitation (details in Sec.~\ref{subsec:photacoustic}).
\end{enumerate}

\begin{figure*}
    \includegraphics[width=\linewidth,keepaspectratio]{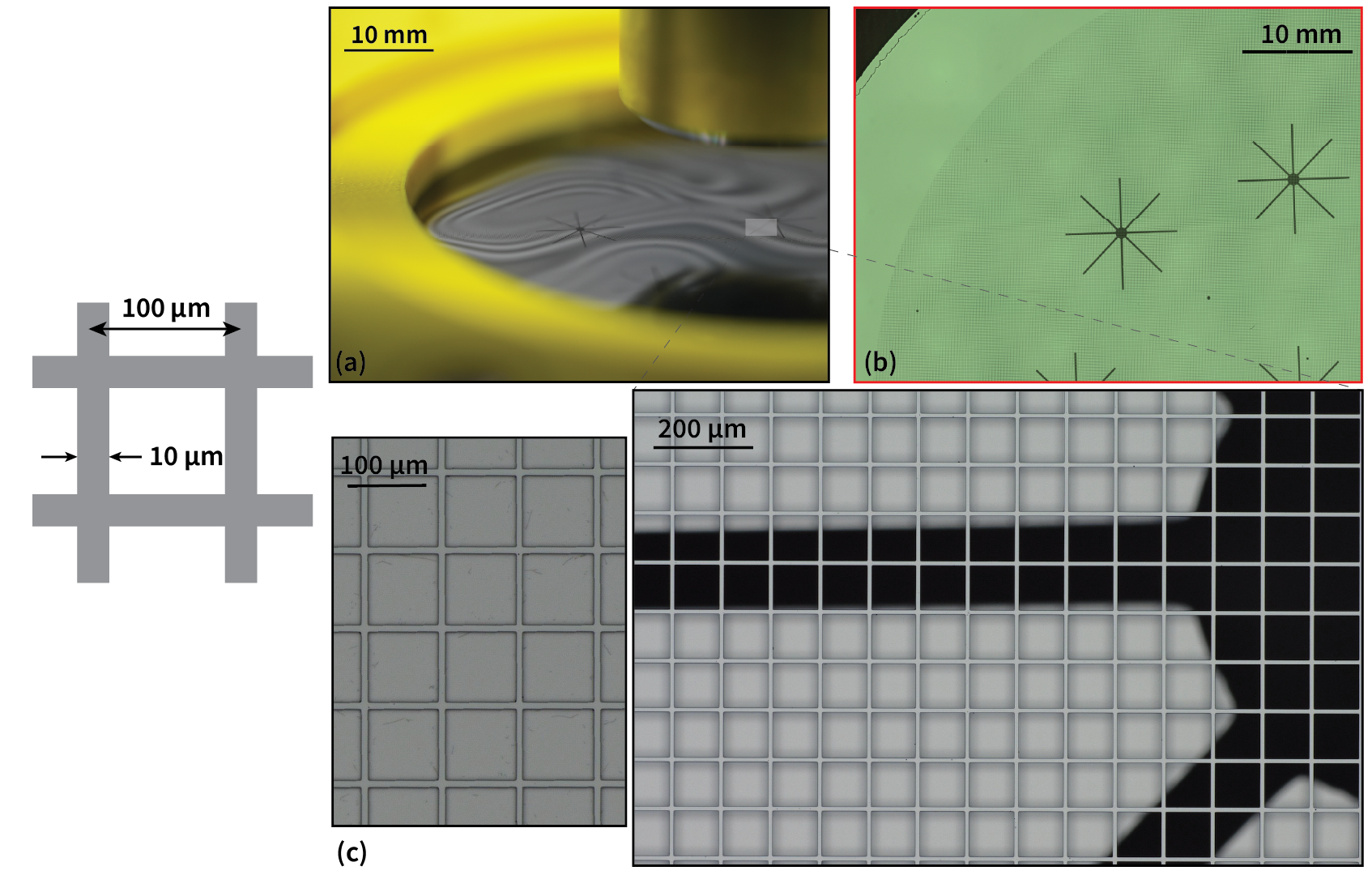}
    \caption{Schematic (left) and optical micrographs of a free-standing periodic microarchitected waveguide with $\sim$600,000 unit cells of dimensions 100~\textmu m and beam width 10~\textmu m within a SOI wafer of 100~mm  diameter. The star-shaped patterns in the micrographs are windows etched in the back (handle layer) for optical access for dynamic pump-probe characterization.}
    \label{fig:fabmicro}
\end{figure*}

The design files for the photo masks (.gds file formats used for the integrated chip layout) were custom built; for periodic architectures, the design files were generated using the open-source software KLayout \cite{klayout}. As this is too tedious for spatially-variant architectures, we developed a Python code, which takes nodal positions and connectivity data of a beam-based computational design as input and writes this data into a .gds file, which incorporates the thicknesses of the elements and nodes for the final mask design. These files are fed as input to the mask writer. This workflow enables the direct passing of information from computational simulations to microfabrication---towards the vision of a closed loop between high-resolution experiment and computational design. Our fabrication protocol overcomes the three challenges presented in Sec.~\ref{sec:intro}: 
\begin{itemize}[leftmargin=*]
    \item Owing to silicon exhibiting very low damping, wave dispersion due to the structural architecture and the underlying base material are effectively decoupled: the observed wave guidance and attenuation stems primarily from the structural design. This enables experimental insight into wave attenuation due to architecture alone, since the base material is nearly non-dissipative. 
    \item The length scales of the unit cell and the free-standing membrane differ by three orders of magnitude, allowing for the study of wave propagation across a wide range of wavelengths relative to the size of the unit cell (broadband characterization), without interference from boundary conditions.
    \item Free-standing membranes enable the investigation of multiple wave modes beyond surface acoustic waves (SAWs).
\end{itemize}

To exemplify this microfabrication procedure, Fig.~\ref{fig:fabmicro} shows a periodic beam lattice with square unit cells, which are illustrated in Fig. \ref{fig:fabmicro}a. Samples were 15~\textmu m-thick (for periodic architectures, Sec.~\ref{subsec:periodic}) and 10~\textmu m-thick (for spatially-graded architectures, Sec.~\ref{subsec:spatialgrading}) free-standing films with maximum diameters of 80~mm, 100 \textmu m unit cells, and minimum beam widths of 5 \textmu m. These ``thin-plate'' samples contain up to approximately 600,000 unit cells, the largest free-standing linear elastic metamaterial waveguides that have been fabricated, in terms of the number and density of unit cells. Note that this is not the limiting resolution, with optical lithography being able to achieve a resolution close to 1~\textmu m, and electron lithography down to tens of~nanometers. The latter, however, is challenging in terms of scalability to entire wafers. 

Optical micrographs of a sample at different length scales are shown in Fig.~\ref{fig:fabmicro}(b,c). Such high-precision samples are fabricated in house within two days. This marks a significant improvement over conventional machining of macroscale metamaterials \cite{dorn2023a,telgen2024rainbow} in terms of scalability ($\mathcal{O}(10^{5} - 10^{6})$ vs.\ $\mathcal{O}(10^{3} - 10^{4})$ unit cells). The excitation and measurement of waves in these \textmu-architected samples pose new challenges, which will be addressed in the following section.

\subsection{Pump-probe photoacoustic characterization \label{subsec:photacoustic}}
\begin{figure*}
    \includegraphics[width=\linewidth,keepaspectratio]{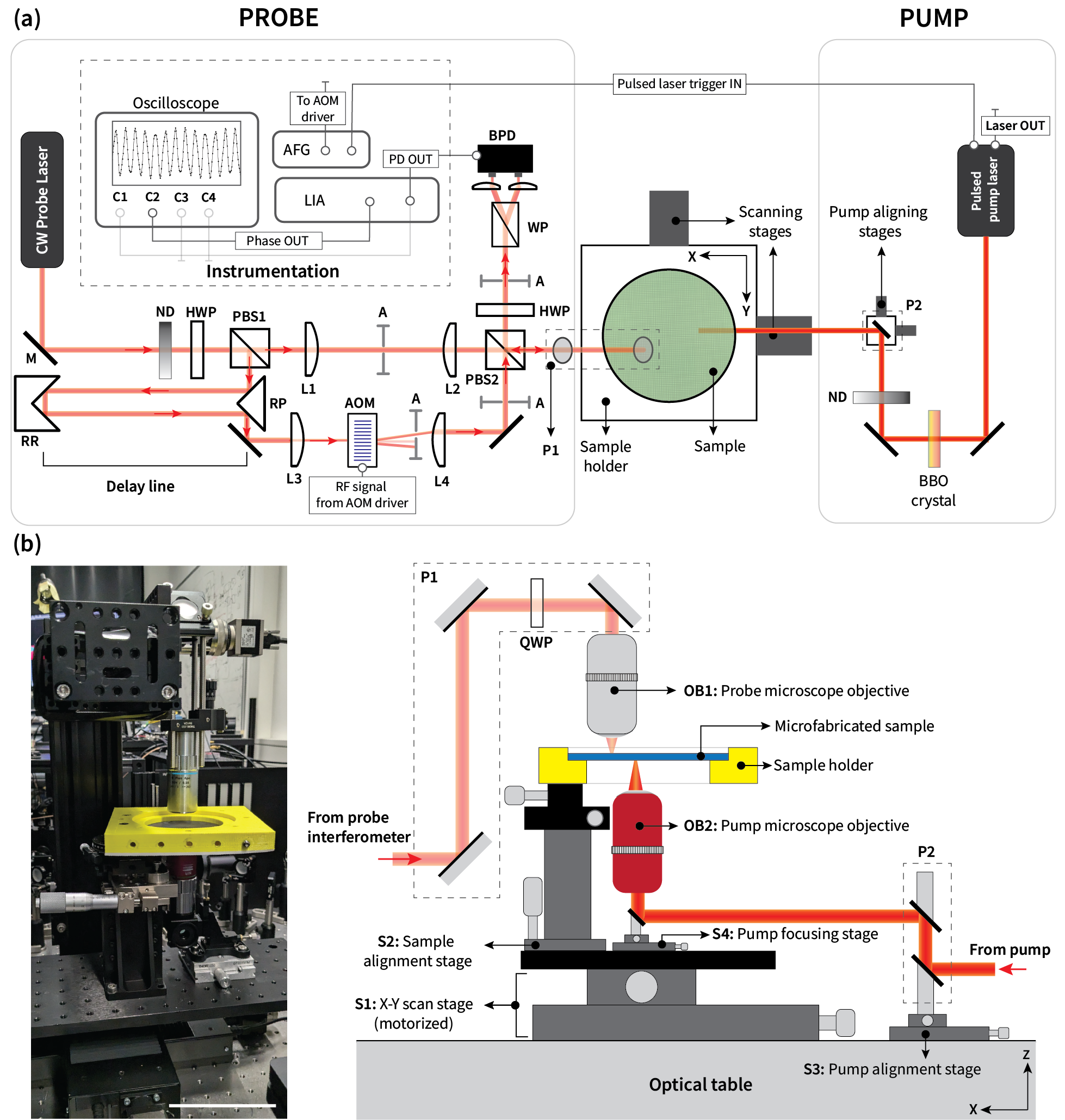}
    \caption{Overview of the pump-probe experiment. (a)~Top view schematic. The section to the left of the sample holder is the probe section -- a home-built heterodyne interferometer with the associated electronic components. To the right of the sample holder is the pump section of the experiment, involving a pulsed laser source with optical elements for filtering and alignment. (b)~Side view. The photograph on the left shows the sample assembly (in yellow) with the Mitutoyo objective on the top to focus the probe laser beam to the point of measurement, and a second objective at the bottom to focus the pump beam (scale bar:~100~mm). Sample and pump objective are mounted on a series of linear stages for alignment and scanning control. The schematic on the right shows details of the optical path, scanning and alignment stages. Acronyms: M - mirror; ND - neutral density filter; HWP - half-wave plate; QWP - quarter-wave plate; PBS - polarizing beam splitter; RP - reflecting prism; RR - retro-reflector; L - plano-convex lens; AOM - acousto-optic modulator; A - aperture; WP - Wollaston prism; BPD - balanced photodetector; P - periscope assembly; AFG - arbitrary function generator; LIA - lock in amplifier.}
    \label{fig:pumpprobe}
\end{figure*}

A photoacoustic pump-probe experiment was developed to resolve elastic wave propagation in our fabricated metamaterials. Fig.~\ref{fig:pumpprobe} shows a schematic of the experiment. The section to the right of the sample is the \textit{pump section} involving a pulsed laser source (see Sec.~\ref{subsubsec:pump}), while the section to the left of the sample is the \textit{probe interferometer} to measure particle velocities resolved in space and time (discussed in Sec.~\ref{subsubsec:probe}). Sec.~\ref{subsubsec:integrated} describes the integration and control of the pump and probe with the sample assembly during a scanning measurement.

\subsubsection{Photoacoustic pump excitation \label{subsubsec:pump}}
Elastic waves were excited using an infrared pulsed laser source~(Coherent FLARE-NX, wavelength 1030~nm, pulse energy 500~\textmu J, pulse width 1~ns). The mechanism of photo-acoustic excitation involves rapid thermal expansion of the thin aluminum film transducer (deposited on both surfaces of the sample) due to an incident laser pulse, generating an acoustic pulse \cite{Krishnaswamy2008,Davies1993}. The interface between the aluminum film and the sample is assumed to be perfect, which enables transmission of the acoustic wave into the sample. 

The pulsed laser is reflected onto the sample through a series of mirrors. The beam passes through a commercially-procured barium borate (BBO) crystal (Fig.~\ref{fig:pumpprobe}(a)) which, when aligned, generates the second harmonic of the infrared laser. This serves the dual purpose of safety and visibility during alignment as well as fine control of the pump energy density to limit ablation of fragile samples. The BBO crystal's use was limited to alignment for this study. The beam also passes through a variable neutral density filter (ND) to control the intensity of the incident pulse. The pump beam then passes through a periscope assembly (P2 in Figs.~\ref{fig:pumpprobe}(a,b)) mounted on an $x$-$y$-stage (S3 in Fig.~\ref{fig:pumpprobe}(b)) to align the pump beam with the sample assembly, and focused on the sample using a 20$\times$ objective lens (OB1 in Fig.~\ref{fig:pumpprobe}(b)). The inset in Fig.~\ref{fig:pumpprobe}(b) shows a photograph of the sample assembly with the pump and probe microscope objectives below and above the sample, respectively. 

All experiments presented in this study were performed using a focused Gaussian laser beam. Careful control and characterization of the spatial and temporal profile of the excitation pulse are feasible for a wide range of applications, e.g., spatial profiling using a micro-lens array (to generate a uniform spatial profile, as opposed to a Gaussian), excitation of the acoustic wave within a narrow wavelength range (for dispersion characterization) \cite{VegaFlick2015,Nelson1982}, generating different spatial distributions of the incident wave, e.g., Gaussian vs.\ line excitations for directional wave propagation \cite{Arias2003}. These are beyond the scope of the current study but are available modifications, which enhance the versatility of the experiment to investigate a wider range of dynamic phenomena.

\subsubsection{Probe:~A custom-built heterodyne interferometer \label{subsubsec:probe}} 
Commercial scanning laser vibrometers have proven to be powerful in studying waves in metamaterials \cite{Ganesh2017,Celli2019,Miniaci2018,telgen2024rainbow}. However, these systems are ineffective at short length and time scales relevant to microarchitected samples. Some of their primary limitations are the range of accessible frequencies, focusing spot size  and hence spatial resolution, the trade-off between scanning resolution and field of view, the signal-to-noise ratio for high-sensitivity measurements, and of course procurement costs. We therefore developed a polarization-sensitive heterodyne interferometer (see the schematics in Fig.~\ref{fig:pumpprobe}) to measure spatially resolved particle displacements from sub-unit cell (\textmu m) to structural (mm) length scales, with a displacement resolution on the order of 1~nm, across time scales spanning tens of ns to ms. 

The heterodyne interferometer is based on the interference of two laser beams --- a reference beam, which is frequency-shifted by a predefined ``carrier'' frequency $\omega_{c}$, and a sample beam reflected from the measurement point \cite{Dewhurst1999}. The intensity of the interfered beams can be written as the intensity of an electromagnetic wave:
\begin{equation}
    \begin{aligned}
       I = ~ & I_{r} + I_{s}
       + 2 I_{r} I_{s} \cos {\Bigg( \omega_{c} t + \dfrac{2 \pi}{\lambda} \delta(t) - \varphi_{0} \Bigg)},
       \label{eq:interference1}
    \end{aligned}
\end{equation}
where 
$I$ is the intensity of an electromagnetic wave, and $\lambda$ its wavelength. 
$I_{r}$ and $I_{s}$ are the intensities of the reference and sample beams, respectively, $\delta(t)$ is the time-resolved out-of-plane particle displacement at the measurement point, and $\varphi_{0}$ is the phase difference between the reference and sample beams under ambient conditions (i.e., when $\delta=0$). The third term in Eq.~\ref{eq:interference1} shows that the measured intensity $I$ is modulated by the carrier frequency~$\omega_c$ under the aforementioned ambient conditions. Particle displacements due to wave motion cause modulation of the phase of this term, hence it is only the argument of this oscillating component (at the carrier frequency $\omega_{c}$) that enables its measurement. The reference and sample intensities, $I_{r}$ and $I_{s}$, respectively, can be controlled through the polarization of the reference and signal laser branches of the interferometer, such that the ambient signal (prior to wave propagation) measured at the oscilloscope is a time-harmonic signal at the carrier frequency, centered at zero. During wave propagation, the particle displacement $\delta(t)$ is directly proportional to the change in phase $\varphi_{m}$ of this signal relative to the ambient reference phase $\varphi_{0}$, such that
\begin{equation}
    \delta(t) = \dfrac{\lambda}{2 \pi} \varphi_{m}.
\end{equation}

Based on the operating principle described above, an interferometer was built with a continuous wave (CW) He-Ne laser (ThorLabs HNL210LB, 20 mW, 633~nm) as the source. Beam intensity is controlled by a variable neutral density filter (ND). The beam first passes through a half-wave plate (HWP) installed before the polarizing beam splitter (PBS1 in Fig.~\ref{fig:pumpprobe}(a)). The HWP is carefully aligned to ensure maximum intensity of the interfered beam, thus increasing the signal-to-noise ratio of the carrier signal (third term in Eq.~\ref{eq:interference1}). The beam splitter PBS1 splits the beam into the reference and signal branches of the interferometer, with orthogonal polarization relative to each other. The latter ensures minimum back-reflection into the laser head, thus minimizing power fluctuations. The reference branch first passes through a delay line, using a reflecting prism (RP) and a retroreflector (RR). The delay line was installed to ensure that the path length difference of the interferometer is restricted to below the coherence length of the laser source (on the order of 10~cm for He-Ne lasers). A crucial component of the reference branch is an acousto-optic modulator (AOM), which shifts the optical laser frequency by 80~MHz. The AOM~(EQ Photonics GmbH, model\#~3080-120) was driven by an RF signal generator~(Gooch \& Housego, model\#~3910), which was triggered by an arbitrary function generator~(AFG -- Teledyne T3AFG40). An aperture was used to selectively transmit the first-order diffracted output from the AOM, frequency-shifted by 80~MHz (carrier frequency).

The signal branch was collimated using two plano-convex lenses~(L1 and L2 in Fig.~\ref{fig:pumpprobe}(a)) and directed to the sample using a periscope assembly (P1). This assembly is shown in the side view schematic (Fig.~\ref{fig:pumpprobe}(b)). After passing through this mirror assembly, the signal beam is focused on the sample through an objective (OB1 in Fig.~\ref{fig:pumpprobe}(b)). The choice of the objective depends on the smallest feature size of the sample---for the experiments in the current study, 20$\times$ or 50$\times$ objectives were used to focus the beam down to spot sizes smaller than 10~\textmu m (the smallest feature size in our samples). The beam was reflected back on the sample path and recombined with the reference branch at the polarizing beam splitter PBS2 (Fig.~\ref{fig:pumpprobe}(a)). The back-reflected signal beam requires its polarization to be rotated by 90$^\circ$ for recombination at PBS2, which is achieved by a quarter-wave plate (QWP in Fig.~\ref{fig:pumpprobe}(b)).

The recombined beams were transmitted through a second HWP and a Wollaston Prism~(WP) and focused on two inputs of a balanced photodetector (BPD; ThorLabs PDB230A, bandwidth 100~MHz). The HWP and WP ensure that signals of equal intensity are directed into the two BPD ports. The BPD output voltage is a differential measure and hence directly related to the difference between the photocurrent from each of the input detectors. This enhances the signal-to-noise ratio, as it eliminates common electronic and optical noise in the system. Careful elimination of noise sources (electronic, optical, and mechanical) is critical for our measurements because of nm-scale displacements during experiments. As we exploit the reproducibility of pump-probe experiments to measure wave motion by sampling at different times after excitation, thousands of excitation and measurement cycles are required, with excitation occurring at the same point. Hence, the pump amplitudes were restricted to minimize sample damage at such high repetition rates. BPD data were analyzed electronically using a lock-in amplifier LIA (Zurich Instruments GHFLI). The lock-in amplifier analyzes the phase of the carrier signal, which is recorded as a function of time by a digital oscilloscope~(Tektronix MSO64B). 

\subsubsection{Sample assembly and integration \label{subsubsec:integrated}} 
The fabricated sample was mounted in a custom-designed, 3D-printed holder (see Fig.~\ref{fig:pumpprobe}(b)). This assembly was mounted on a composite stack of translation stage assemblies S1 and S2 (Fig.~\ref{fig:pumpprobe}(b)) to enable independent alignment and motion of the sample with respect to the pump and probe. S1 comprises two motorized translation stages (ThorLabs NRT150) mounted orthogonal to each other for independent translation in the $x$ and $y$ directions (see the definition of the coordinate axes in Fig.~\ref{fig:pumpprobe}(a)). The sample assembly and pump focus stage (S4) are mounted on S1, controlled remotely using a Python code. This enables automated scanning measurements along predefined paths in the $x$-$y$-plane, while ensuring repeated pump excitation at the same point on the sample. The stage assembly S2 is used for manual alignment of the sample assembly with respect to the pump and/or probe beams, with two translational degrees of freedom along the $y$- and $z$- directions, and one rotational degree of freedom about the $z$-axis. S3 controls the alignment of the pump beam from the optical table to the sample (allowing free choice of the excitation location). The current configuration restricts each automated experiment to line scans parallel to the $x$-direction (Fig.~\ref{fig:pumpprobe}(b)), with multiple line scans required for full 2D reconstruction of a propagating wave (demonstrated in Sec.~\ref{subsec:spatialgrading}). Therefore, careful alignment of the pump beam with the S1 scanning stage assembly is necessary to ensure repeatable excitation during each automated line scan; this is achieved using the pump alignment stages S3 and S4 (Fig.~\ref{fig:pumpprobe}(b)).  Future upgrades, involving coupling the free-space pump laser to optical fibers, can enable completely arbitrary area scans in a single automated experiment. 

Before each line scan, stages S2, S3, and S4 were carefully aligned for optimal excitation conditions and the best signal-to-noise ratio of the probe carrier signal. Each line scan was performed remotely, using a Python code (see Appendix~\ref{appsec:control} for descriptions and timing diagrams). 

\section{Results and discussion \label{sec:results}}
We first present proof-of-concept results from the new experiment applied to periodic architectures. Comparison with data from prior studies by the authors at macroscopic length scales shows excellent agreement and repeatability. We furthermore demonstrate the versatility of the experiment to realize novel, non-periodic waveguide designs that have been studied only computationally to date.

\begin{figure*}
    \includegraphics[width=\linewidth,keepaspectratio]{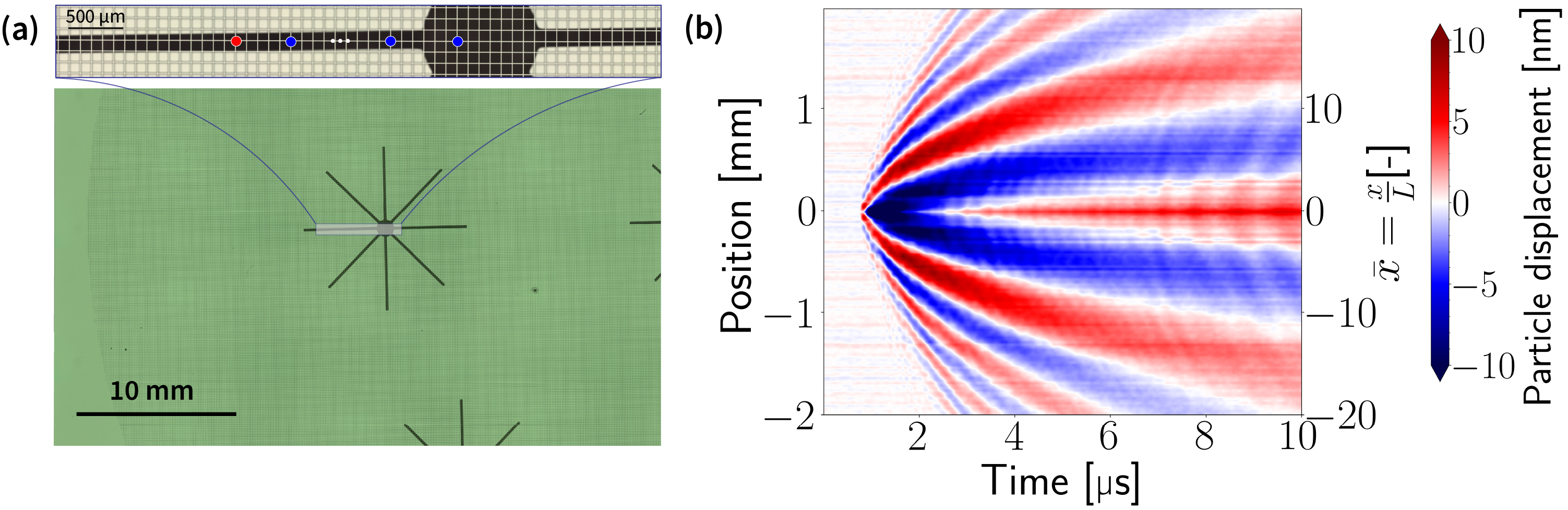}
    \caption{Wave propagation in a periodic film. (a)~Schematic of the experimental scanning protocol superimposed over an optical micrograph of a periodically architected sample with square unit cells (same dimensions as in Fig.~\ref{fig:fabmicro}) with an inset showing scanning locations. A line scan was performed with pump excitation at the red circle (see inset), and probes along the line corresponding to the blue circles. A scanning resolution of $\sim 25$~\textmu m was used for this experiment. (b)~Reconstruction of the position-time data from a single scanning experiment. Raw data was plotted after windowing and interpolated using a nearest-neighbor bilinear fit.}
    \label{fig:periodic_schema}
\end{figure*}

\subsection{Elastic waves in periodic architectures \label{subsec:periodic}}
Elastic wave guiding through periodic architectures has been studied extensively at the macroscopic length scales (order of cm). For example, Telgen et al.~\cite{Telgen2024} recently explored the dispersion relations of such architectures both experimentally and numerically. We test our fabrication and characterization protocol on similar periodic architectures, here fabricated with at least two-orders-of-magnitude improvement in spatial resolution and number of unit cells. 

A schematic of the experimental protocol is shown in Fig.~\ref{fig:periodic_schema}(a). Pump pulses were generated with rise times on the order of 10-100~ns and durations of $\sim$1~\textmu s at the point marked by the red circle in Fig.~\ref{fig:periodic_schema}(a) (see inset). The variation of rise times relates to pulse-sample interactions, requiring careful control. Following each pump excitation, the sample stage was translated by 25~\textmu m (resulting in a lateral resolution of four points per unit cell) to collect out-of-plane displacement data from each spatial point. This procedure was repeated for 80 points along the line marked by the blue circles in Fig.~\ref{fig:periodic_schema}(a). Before each experiment, the pump optical path was aligned to ensure no observable change in excitation location throughout the line scan. This was verified by observation on a charge-coupled device (ccd) camera integrated into the optical line of the interferometer.  

The raw experimental data captures a wide range of wave types, including long-wavelength modes representing waves and vibrations relevant to the plate rather than the microarchitecture. Therefore, datasets were analyzed and post processed using a Python code (see Appendix~\ref{appsec:analysis}). Each time-resolved signal was cropped to represent the time of interest for short-wavelength waves (approaching the length scale of the unit cell). Data were windowed in time using a Tukey window to minimize spurious oscillations in Fourier space due to the finite duration of the signal. The windowed data is illustrated on a 2D Lagrangian position-time diagram in Fig.~\ref{fig:periodic_schema}(b). Note that the data is mirrored about the zero position for ease of Fourier analysis in space. This is valid due to the periodicity of the structural design. As expected, the iso-contours do not form straight lines, indicating dispersion of the wave as it propagates through the unit cells. 

\begin{figure*}
    \includegraphics[width=\linewidth,keepaspectratio]{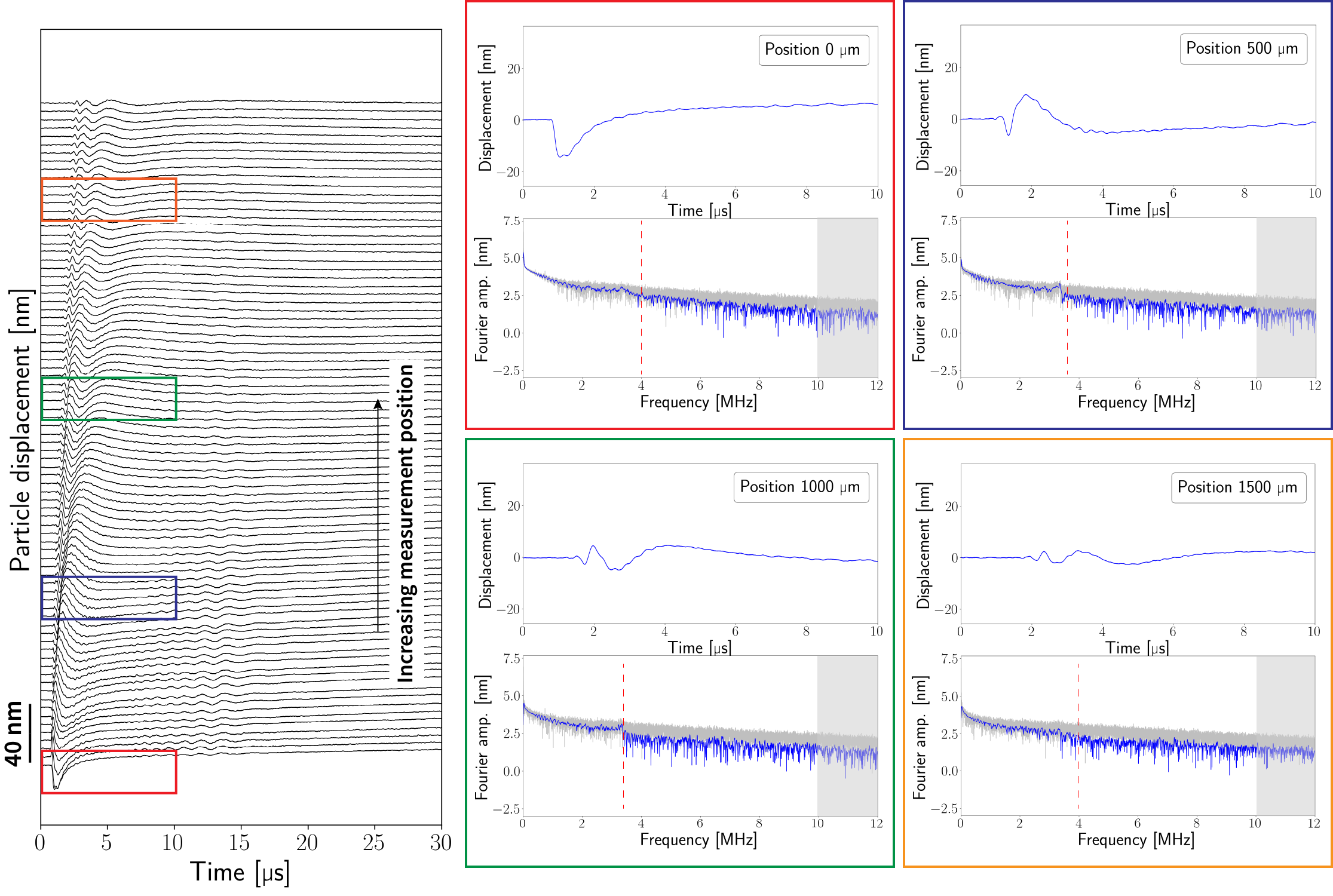}
    \caption{Windowed displacement-time data for all measurement points of the experiment on a periodic square unit-cell architecture. On the left, the data at each spatial location is offset along the $y$-axis for visualization. The scale bar on the left indicates the amplitude of each time resolved signal. Images on the right (with matching colors) show selected raw and windowed time-resolved data and the corresponding Fourier spectra. Gray spectra correspond to raw data, while the blue ones are windowed data. Dashed lines indicate the maximum frequency where changes in the Fourier spectra relative to excitation are observed. Data cannot be resolved beyond 10~MHz due to the physical limitation of electronic demodulation (marked by the gray region).}
    \label{fig:periodic_time}
\end{figure*}

Alternative representations are shown in Fig.~\ref{fig:periodic_time}. The plot on the left shows the particle displacement of each measurement point (offset along the $y$-axis for visualization), as a function of time. Four representative data sets (indicated by the four colors) are shown with more details on the right. The latter compare the windowed particle displacement data along with the Fourier spectra of the raw data (in gray) and the windowed data (in blue). The excitation signal (data at position 0~\textmu m) shows a rise time on the order of 100~ns. At different positions, Fourier data show qualitative changes (relative to the excitation signal) up to $\sim 4$~MHz (indicated as dashed red lines in the Fourier spectra). This is noticeable at positions $500$ and $1000$~\textmu m. Although the displacement resolution is $<1$~nm, data beyond 10~MHz is not measurable using the current configuration of the interferometer and its electronics (indicated by the gray region in the Fourier spectra). Improved characterization and control of the resolution limits of the interferometer for future high-precision measurements are underway.

\begin{figure*}
    \includegraphics[width=\linewidth,keepaspectratio]{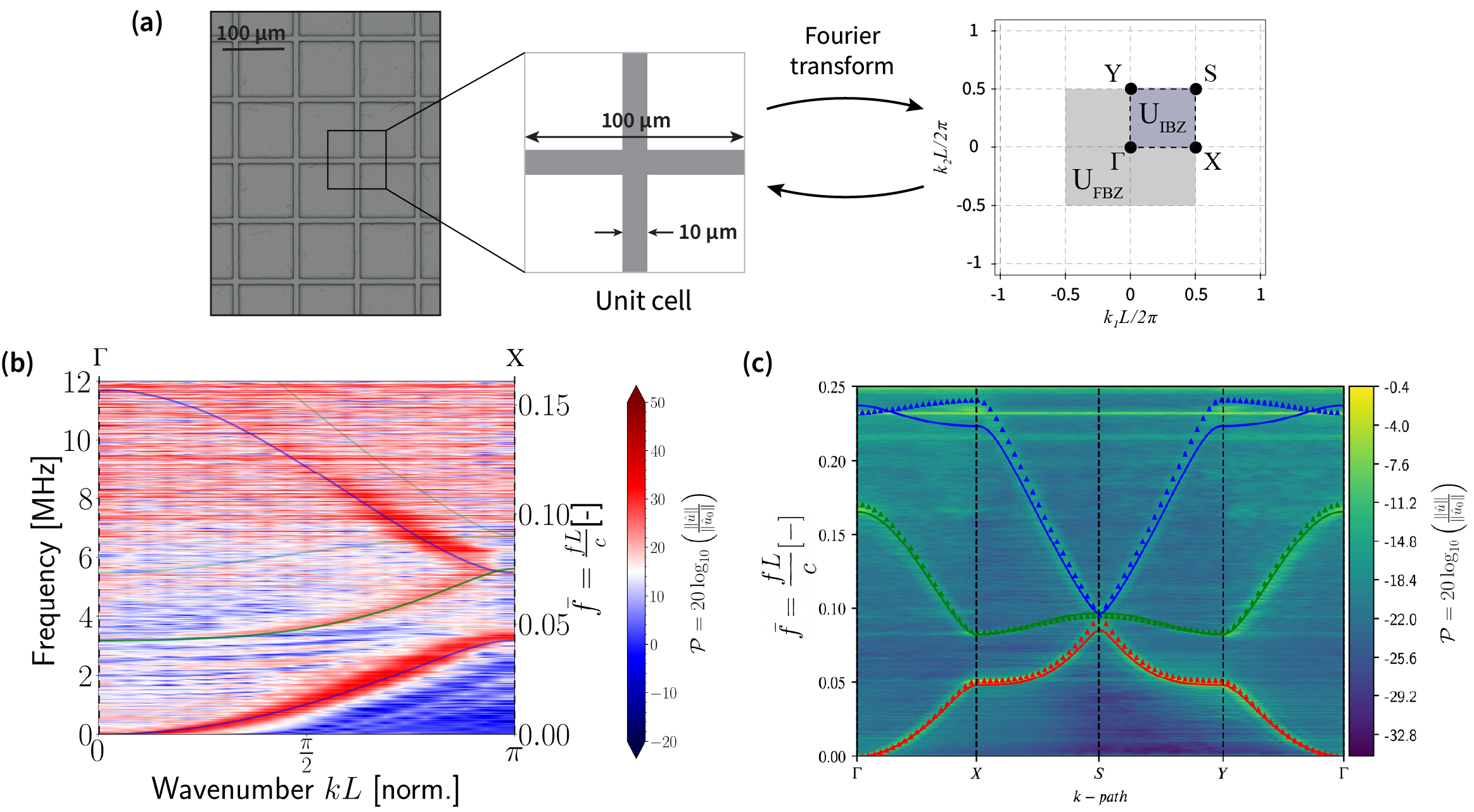}
    \caption{(a)~Experimental micrograph with a magnified view of the unit cell, and a schematic of the first and irreducible Brillouin zones, denoted as $U_{FBZ}$ and $U_{IBZ}$, respectively. Data in (b) and (c) are plotted along sections of the symmetry path $\Gamma-X-S-Y$ for comparison between the micro- and macro-scale wave guides. (b)~Wave dispersion through periodic microarchitecture. The color map is the experimentally-constructed section of dispersion surfaces, while the solid lines are the dispersion relations obtained from fully-resolved finite element calculations. (b)~Data from Telgen et al.~\cite{Telgen2024} on ``macro-architectures'' (i.e., metallic samples of millimeter-scale unit cells). The $\Gamma-X$ section of the dispersion plot in (c) is equivalent to the full plot in (a). The latter does not show the full $\Gamma-X-S-Y$ path as in (b) due to scanning along only one spatial direction. Excellent match between microscale (a) and macroscale (b) dispersion plots validate of our microscale experimental paradigm.}
    \label{fig:periodic_freq}
\end{figure*}

To extract wave dispersion information from the data, a 2D Fourier transform was performed on the spatio-temporal data set (represented in Figs.~\ref{fig:periodic_schema}(b) and \ref{fig:periodic_time}). The Fourier spectrum of each time-resolved signal $\hat{u}(f;x=x_{i})$ was normalized against that of the excitation point $\hat{u}_{0} = \hat{u}(f;x=x_{0})$. The resulting data shows only a section of the 2D Brillouin zone due to scanning only along one spatial direction. Multiple line scans were not necessary for this dataset due to the periodicity of the architecture implying symmetry of the Brillouin zone (schematically shown in Fig.~\ref{fig:periodic_freq}(a)). Fig.~\ref{fig:periodic_freq}(b) presents a color map of the thus-obtained data in the frequency ($f$) vs.\ (normalized) wave number ($k$) space. The wave number is normalized by the length of a unit cell, $L = 100$~\textmu m. The vertical axis on the right indicates the normalized frequency $\bar{f} = \dfrac{fL}{c}$ after \cite{zelhofer2017acoustic,Telgen2024}, where $c \approx 8500$ m/s is the longitudinal wave speed of the material (in this case single-crystal silicon \cite{hopcroft2010young}). These normalized metrics allow for a direct comparison of wave dispersion in structures across length and time scales. The solid lines represent dispersion relation data from fully resolved finite element calculations (see Appendix~\ref{appsec:fe} for simulation details).

Fig.~\ref{fig:periodic_freq}(c) from Telgen et al.~\cite{Telgen2024} shows equivalent data for a macroscale metamaterial (dimensions $40 \times 40$~cm, unit cell size $5$~mm, beam thickness $500$~\textmu m) machined from aluminum. In their experiments, full 2D data were collected, hence probing $k$-space along two directions. The range of the horizontal axis in Fig.~\ref{fig:periodic_freq}(b) is equivalent to the path $\Gamma-X$ in Fig.~\ref{fig:periodic_freq}(c). Lines and symbols represent computed dispersion relations from finite element calculations (referring to beam and solid finite elements, respectively).

Data from our new microarchitected wave guides (Fig.~\ref{fig:periodic_freq}(b)) shows modes that are resolvable up to $\sim 10$~MHz, i.e., up to a normalized frequency $\bar{f} \simeq 0.15$. In agreement with the macroscale experiments of~\cite{Telgen2024} (Fig.~\ref{fig:periodic_freq}(c)), this corresponds to the range covering the first and second out-of-plane modes. The major limiting factor towards resolving higher-frequency modes is the minimum time constant of the lock-in amplifier (14~ns). This limit can be pushed through experimental design, but this falls beyond the scope of the current study. 

The match between finite element results and microscale experiments is excellent (see Fig.~\ref{fig:periodic_freq}(b)), showing the accuracy of the new microscale wave guides and the pump-probe experiment. However, the data shows an additional intermediate mode between the expected modes. In comparison with finite element data (green lines in Fig.~\ref{fig:periodic_freq}(b)), this matches a wave mode that is expected to propagate along the $X-S$-path in $k$-space (see Fig.~\ref{fig:periodic_freq}(b)), which is perpendicular to the direction probed in our microscale experiments. Although the exact mechanisms for appearance of this mode are currently unknown, we hypothesize that this could occur either from scattering of waves from defects (e.g., damaged beams or dust particles) in the vicinity of the scan path, or from potential out-of-plane curvature of the microfabricated film. Careful control of the pump pulse and etching parameters is expected to limit these effects.

\subsection{Elastic wave guiding in spatially-graded architectures \label{subsec:spatialgrading}}
\begin{figure*}
    \includegraphics[width=\linewidth,keepaspectratio]{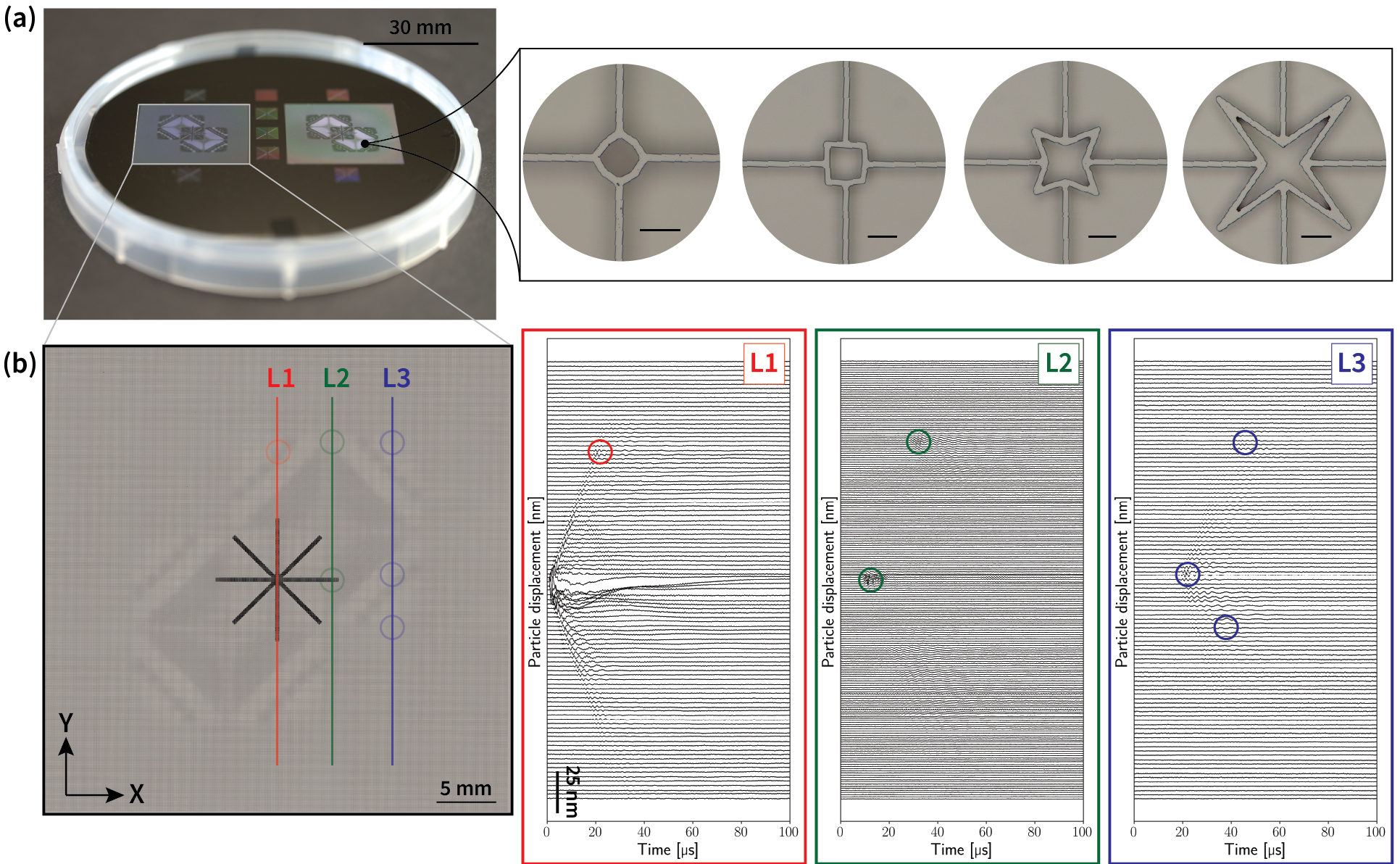}
    \caption{Wave guiding through a spatially-graded microarchitected film. (a)~Picture of the full microfabricated wafer. For redundancy, two ``samples'' were fabricated on a single wafer. Insets show individual unit cells (scale bar: 25~\textmu m) -- the architecture was uniformly graded across these unit cell designs. (b)~Optical micrograph of one of our sample devices. Three line scans were collected along lines L1, L2 and L3. Raw displacement-time data plotted on the right of (and aligned with) the micrograph show evidence of the wave directed along a figure-8 pattern.}
    \label{fig:graded}
\end{figure*}
Following the successful test of the new fabrication-experimental protocol on periodic designs, we now present its ability to fabricate spatially-graded waveguide architectures and characterize their wave motion in situ. Although spatial grading has proven promising for wave guiding at larger length scales (see, e.g., \cite{Telgen2024, dorn2023a}), those samples lacked a clear separation of scales between the unit cell dimensions and the structural length scale of the samples, implying limitations on ability to generate arbitrarily smooth gradients. This has limited our ability to design graded waveguides. The microfabrication technique developed in this study (Sec.~\ref{subsec:fabrication}) overcomes that limitation. Fig.~\ref{fig:graded} shows an example of a spatially graded film. whose architecture was obtained from an inverse design scheme~\cite{Dorn2023b} to guide waves in a ``figure-8'' pattern. Two independent wave guides were fabricated on a single SOI wafer (see the two large regions in Fig.~\ref{fig:graded}(a), each referred to as a ``sample''), since the design did not require the use of an entire 100~mm wafer. This is yet another advantage of the developed experimental protocol: arbitrary control of the number of waveguides fabricated on a single wafer, based on applications of interest. The smaller architected regions on the wafer (referred to as ``test areas'') were used to optimize the intensities and frequencies of the pump and probe beams before the actual experiment. This ensured minimal damage to the pump region of the sample during the actual experiments. 

The inset in Fig.~\ref{fig:graded}(a) shows high-resolution micrographs of individual unit cells in the spatially graded samples (each scale bar 25~\textmu m). The design involves smooth gradients between the shown unit cells, with a  uniform beam thickness of $5$~\textmu m and unit cell sizes of $100$~\textmu m. The choice of the beam diameter had two reasons: (1)~to test the limits of our fabrication protocol, and (2)~to maintain the assumption of slender beams (which admits the accurate prediction of dispersion relations via beam-based finite elements; see a detailed discussion and quantitative comparison in \cite{Telgen2024}). Fig.~\ref{fig:graded}(b) shows a micrograph of one spatially graded sample. As mentioned in Sec.~\ref{subsec:photacoustic}, the back-window was used to excite the acoustic wave at the center of the design for all measurements. Three line scans, labeled L1, L2, and L3, were performed. Arrival and arrest locations of the signal are marked by circles in Fig.~\ref{fig:graded}(b). At each measurement point, 50~time series datasets were averaged to increase the signal-to-noise ratio. The corresponding particle displacement data is shown in Fig.~\ref{fig:graded}(b), where the positions on the vertical axis of all time series data correspond to the vertical positions in the micrograph on the left. Scanning resolution varied between $200$~\textmu m (for L1 and L3) and $100$~\textmu m (for L2). All measurements were taken for a total line length of 20~mm, hence spanning 200~unit cells. Note that the particle displacements in these experiments are much smaller than those in the experiments on periodic samples reported above. This is partly due to a lower pump pulse energy (used to minimize ablation of the aluminum coating during long-time measurements) and the inherent wave guiding ability of the architecture.

Line scan L1 shows relatively large displacements on the order of $10$~nm at the excitation point. The wave propagates along L1 at a group velocity of approximately 275~m/s (approximated from the average slope in the position-time diagram). We notice a gradual drop in signal (more than an order of magnitude) over $\sim 6$~mm (i.e., 60~unit cells away from the excitation point -- marked by the red circle in plot L1 of Fig.~\ref{fig:graded}(b)), followed by loss of signal (below the noise floor) after 70~unit cells from the excitation point. The elastic wave does not propagate beyond this range along L1 because the guided wave is redirected in the direction perpendicular to L1 toward L2. 

Line scan L2 shows the acoustic wave arriving at its center $\sim 8$~\textmu s after excitation (implying a wave speed $\simeq 400$~m/s). Notice the anisotropy in the wave speed between the $x$- and $y$-directions. Beyond wave guiding, this also confirms directionality in the transient wave front propagation. The acoustic signature disappears after 2-3~unit cells from the center point and reappears around the 60th~unit cell in the positive $y$-direction (green circle). We do not see this signature in the negative $y$-direction. Line scan L3 shows the wave arriving at $\sim 18$~\textmu s after excitation. At increasing distances from the center, we observe slight signatures of the wave in the positive $y$-direction for up to $60$ unit cells, but only up to $\sim 10$ unit cells in the negative $y$-direction. These observations support the results of the computational design, viz.\ that the excited acoustic wave is indeed guided in a figure-8 pattern centered at the excitation point. (These characteristics are even more evident after numerical bandpass filtering \cite{DornKannan2025}.) Aside from confirming the ability to effectively guide waves in complex spatially-graded structures, this also demonstrates the powerful toolset provided by the introduced microfabrication route for wafer-based metamaterials and the presented in-situ measurement setup.

\section{Conclusion} \label{sec:conclusions}
We have presented a new experimental protocol that enables and characterizes elastic wave guiding in free-standing micro-architected materials. Our fabrication protocol adopts state-of-the-art silicon microfabrication techniques to generate free-standing microarchitected films with demonstrated spatial resolutions down to $5$~\textmu m, and unit cell densities on the order of $10^{3}$ per mm ($\sim 6 \times 10^{5}$ unit cells on a single free-standing film). These samples support a wide range of elastic wave modes across length scales spanning millimeters down to $10$~\textmu m, and time scales of tens of nanometers to hundreds of microseconds. 

The dynamic characterization of these wave modes was performed experimentally by a scanning optical pump-probe experiment involving a pulsed laser source to excite elastic waves and a home-built polarization-sensitive heterodyne interferometer to measure particle displacements. Automated scanning measurements on periodic architectures show excellent agreement with finite element calculations and prior experiments on larger-scale meta-structures, hence validating the presented experimental scheme. To demonstrate the value of our experiment beyond periodic architectures, we developed a spatially graded architecture, which guides an elastic wave along a figure-8 path, based on a computational framework for the inverse design. Excellent agreement between the experimental data and figure-8 design both confirms the applicability of the microfabrication approach to spatially graded architected materials as well as the effectiveness of the measurement setup for its characterization. 

The presented experimental framework opens up new opportunities in high-throughput experimental data-driven discovery of wave dispersion relations, necessary as input for the inverse design of spatially-graded elastic wave guides. In addition to the realization and validation of new designs, this has the potential to break new ground in automated real-time feedback between the experiment and computation towards the discovery of new wave-guiding architectures.  

\begin{acknowledgments}
Mask writing and fabrication were performed in the cleanroom facility of the Binnig and Rohrer Nanotechnology Center of IBM Zurich. The authors thank Dr.~Emil Bronstein for his assistance in setting up experimental scans. C.D.\ acknowledges partial support from an ETH Zurich Postdoctoral Fellowship.
\end{acknowledgments}

\section*{Data Availability Statement}
Data and codes supporting this study are publicly available in a repository accessible at \url{https://doi.org/10.3929/ethz-b-000743304}.

\appendix
\section{Pump-probe experiment control \label{appsec:control}}
\begin{figure}[!h]
    \includegraphics[width=\linewidth,keepaspectratio]{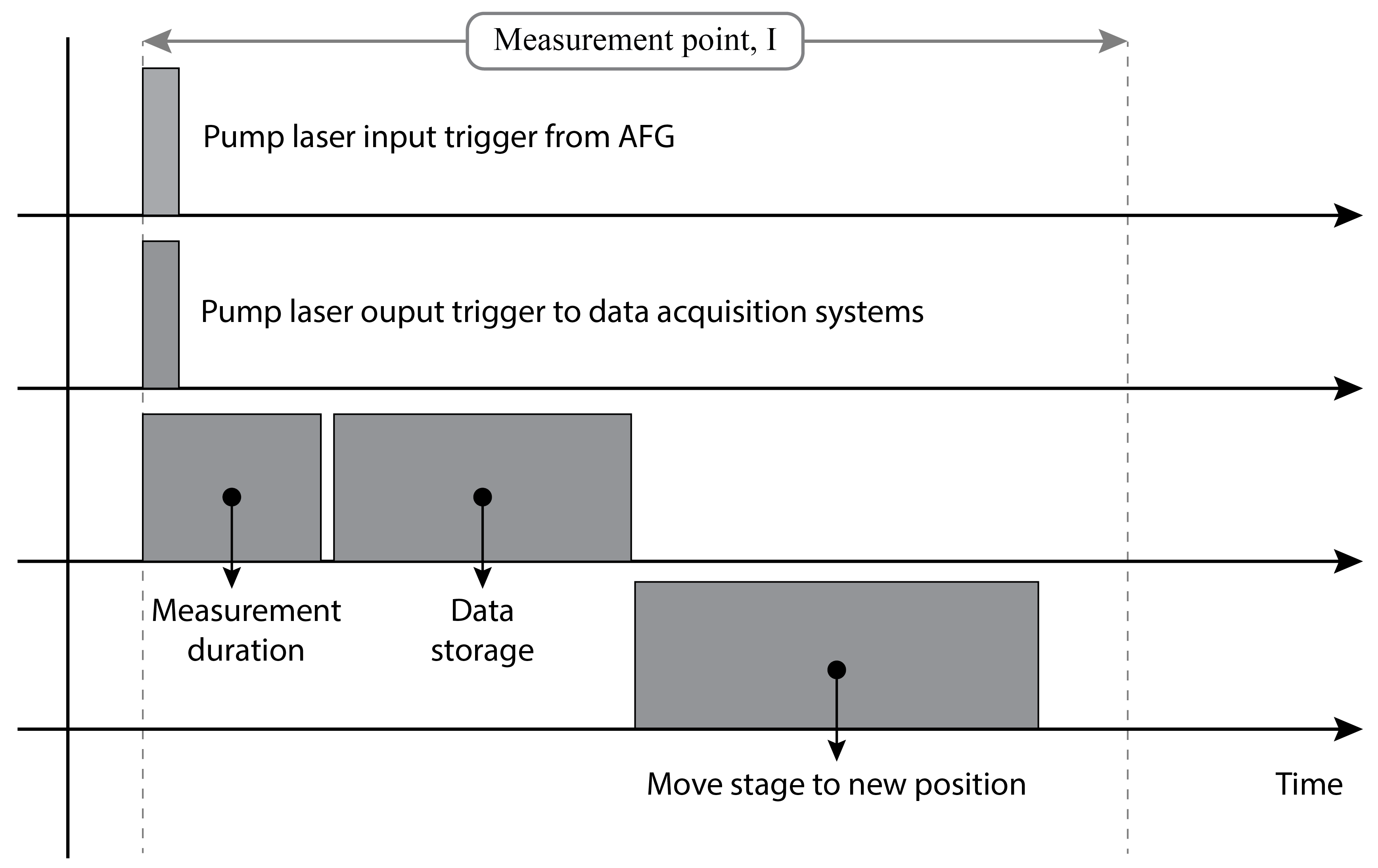}
    \caption{Timing diagram for a single line scan pump-probe experiment.}
    \label{fig:timing}
\end{figure}

Experiments were semi-automated through a combination of instrumentation and control using a home-built Python code. The code controls the input parameters of the oscilloscope and scanning stages. A flow chart of the automation algorithm along with a schematic timing diagram is shown in Fig.~\ref{fig:timing}. Before an experimental scan, the following parameters are fed as input to the automation code: (1)~oscilloscope acquisition and trigger settings, (2)~step size, speed and number of steps of the scanning stages. The lock-in amplifier parameters are also set prior to the experimental scan. One channel of the function generator sends a DC voltage to the AOM driver of the probe interferometer. The second channel of the function generator is programmed to send pulses at a desired frequency to the control unit of the pump laser. The experimental scan and instrumentation are triggered by the output channel of the control unit, so that data acquisition of each measurement point throughout a scan is autonomous and synchronized with the pump pulse. 

At the start of the experiment, the function generator channels are turned on before running the Python code. Pulses from the AFG are read by the pump laser control unit, which sends read-out signals to the oscilloscope trigger input channel upon triggering each optical pulse. Before each acquisition, the input parameters are sent to the oscilloscope, which is then armed for acquisition. After acquisition, the data is stored locally on the hard drive of the oscilloscope, before triggering one step of the scanning stage. Although the duration for this motion lasts on the order of milliseconds, a wait time of 1~s is imposed to ensure that the scanning stages are stabilized before the next acquisition. This then triggers the next acquisition in a loop, until the maximum number of steps entered a-priori as input is reached. The scanning stage motion was verified to be accurate until $5$~\textmu m, beyond which it was unnecessary to test for the purpose of our experiments. However, this is not the limiting scanning resolution achievable for high-resolution measurements. Future versions of the automation code can involve remote control of the AFG trigger to the pump laser, to limit the number of pump pulses on the sample, and real-time storage and analysis of data on a centralized server. The implementation of these modifications is relatively straightforward, with the exception of requiring large storage space and fast data transfer for high-resolution, long-time measurements. However, such a system could advantageously alter the paradigm of experiment-driven computational design.  

\section{Data analysis \label{appsec:analysis}}
Fourier analysis was performed on the raw averaged time series data collected at each measurement point. Each time series signal was first cut within the temporal region of interest. A standard cosine-tapered, also known as Tukey, window was applied to all time-series data to minimize ringing artifacts in the Fourier transform of a truncated signal. The Tukey window parameters were chosen so as to gradually reduce the signal amplitude from maximum to zero, without affecting the signal of interest. Position-time plots (in Fig.~\ref{fig:periodic_time}) were generated using standard plotting tools with bilinear interpolation between adjacent spatio-temporal points. 

\section{Finite element analysis \label{appsec:fe}}

Dispersion relations for the periodic beam lattice were computed numerically for comparison to experimental data. Finite element analysis computed the dispersion relations based on standard formulations presented in, e.g., \cite{phani2006wave, hussein2014dynamics}. We use an in-house finite element code \cite{ae108} for these computations.

The basis for this analysis is a description of the structure by Timoshenko beam finite elements, modeling a single unit cell. Timoshenko beam elements provide an excellent approximation for the lowest out-of-plane modes as compared to higher-fidelity but numerically inefficient solid finite elements; see \cite{Telgen2024} for further discussion. Our mesh exhibits a maximum element size of $L/20$, where $L=100 \ \mu$m is the lattice spacing, which was found to be sufficiently fine to resolve the first two out-of-plane bands accurately.

In the finite element setup, silicon is modeled as a linear elastic material, adopting the elastic properties of \cite{hopcroft2010young}. While silicon exhibits moderate anisotropy, the beams of the square lattice are aligned with the $110$ and $\bar{1}10$ crystal axes of the wafer (see \cite{hopcroft2010young} for orientation definitions) and, since we only consider out-of-plane bending of beams aligned with these axes, we model the material as isotropic with the relevant directional properties of Young's modulus $E=169$~GPa and Poisson's ratio $\nu=0.064$.

For a given wave vector, Bloch boundary conditions are applied to finite element mesh of the unit cell, and the resulting eigenvalue problem is solved to compute the eigenfrequencies and associated mode shapes. The out-of-plane modes are determined by examining the mode shapes, as described in \cite{Telgen2024}. Fig.~\ref{fig:periodic_freq} shows the first two dispersion branches corresponding to the $k_1=0$-axis.

\bibliography{waves_bib}

\end{document}